\documentclass[a4paper]{article}

\usepackage{INTERSPEECH2022}
\usepackage{multirow}
\usepackage{cite}

\title{Unsupervised Uncertainty Measures of Automatic Speech Recognition for Non-intrusive Speech Intelligibility Prediction}
\name{Zehai Tu, Ning Ma, Jon Barker}
\address{University of Sheffield, Department of Computer Science, Sheffield, UK}
\email{\{ztu3, n.ma, j.p.barker\}@sheffield.ac.uk}

\begin{document}

\maketitle
\begin{abstract}
Non-intrusive intelligibility prediction is important for its application in realistic scenarios, where a clean reference signal is difficult to access. The construction of many non-intrusive predictors require either ground truth intelligibility labels or clean reference signals for supervised learning. In this work, we leverage an unsupervised uncertainty estimation method for predicting speech intelligibility, which does not require intelligibility labels or reference signals to train the predictor. Our experiments demonstrate that the uncertainty from state-of-the-art end-to-end automatic speech recognition (ASR) models is highly correlated with speech intelligibility. The proposed method is evaluated on two databases and the results show that the unsupervised uncertainty measures of ASR models are more correlated with speech intelligibility from listening results than the predictions made by widely used intrusive methods.

\end{abstract}

\section{Introduction}
Speech intelligibility is usually interpreted as how comprehensible speech is. Accurate intelligibility prediction has always been of great interest for its importance in developing speech enhancement related applications, such as hearing aids. In recent years, non-intrusive intelligibility prediction, which does not require clean signals as references, has drawn increasing attention because of its wider applicability compared to intrusive methods, especially in realistic scenarios. One of the promising candidates for non-intrusive intelligibility prediction is ASR models~\cite{holube1996speech, jurgens2009microscopic, spille2018predicting, karbasi2020non}, given that they can perform similarly to human speech recognition in certain situations~\cite{barker2007modelling, schadler2015matrix, fontan2017automatic}. 
Intelligibility can be characterised by the probability of correct word recognition~\cite{allen1995humans}. Meanwhile, the \textit{uncertainty} of ASR models is associated with the probability of models making correct predictions~\cite{malinin2020uncertainty, oneactua2021evaluation, kalgaonkar2015estimating, ragni2018confidence, swarup2019improving}.

Motivated by this, this study investigates how to estimate the uncertainty of a strong ASR model and correlate it to speech intelligibility. Specifically, we propose to use an unsupervised ASR uncertainty estimation method, which does not require intelligibility labels or clean references to predict sequence-level speech intelligibility. Our experiments are conducted on both a small vocabulary database with simple noisy scenes and a large vocabulary database with more complex noisy scenes. It is shown that the uncertainty of strong ASR models is highly correlated to speech intelligibility, and the prediction performance can outperform widely used intrusive intelligibility predictors. The experimental results also indicate that the uncertainty of ASR models is better than ASR recognition results at intelligibility prediction.

The next section presents the background for unsupervised ASR uncertainty estimation and recent non-intrusive intelligibility prediction methods. The methodology used to formulate unsupervised ASR uncertainty is explained in Section~\ref{sec:method}. Section~\ref{sec:experiments} describes the experimental data, setups, and results. Section~\ref{sec:conclusion} concludes this work.

\vspace{-1.5mm}
\section{Background}
\label{sec:background}
\vspace{-0.5mm}
Uncertainty estimation is crucial for ASR application as it can help improve robustness in critical tasks. Most ASR uncertainty estimation methods construct and optimise an estimator on top of the original ASR model with supervision~\cite{kalgaonkar2015estimating, ragni2018confidence, swarup2019improving}. Recently, a word-level ASR uncertainty estimation method is proposed in~\cite{oneactua2021evaluation}, and a sequence-level uncertainty estimation method for auto-regressive structured prediction tasks is proposed in~\cite{malinin2020uncertainty}. The major advantages of sequence-level uncertainty estimation for intelligibility prediction, which is used in this work, are that it does not require: firstly, human listening results because they are usually noisy and expensive; secondly, token-level labels because the alignment could be difficult and intractable, i.e., listeners may respond little when the speech is not intelligible.

Conventional non-intrusive intelligibility predictors, such as SRMR~\cite{falk2010non} and ModA~\cite{chen2013predicting}, take advantage of acoustic features related to intelligibility. They heavily rely on prior knowledge on scene acoustics, such as room reverberant characteristics, therefore the application is limited. Another group of methods can be considered as variants of intrusive predictors, like the short-time objective intelligibility\,(STOI)~\cite{taal2011algorithm}, such as NI-STOI~\cite{andersen2017non}, NIC-STOI~\cite{sorensen2017non}, THMMB-STOI~\cite{karbasi2016twin}. A clean feature estimation model is usually constructed and used to produce an estimated reference for computing STOI-like scores. Therefore, clean signals are usually required to optimise the estimation model. Meanwhile, transcription or clean signals are sometimes preferred for alignment or voice activity detection. Recently, a number of data-driven methods are proposed, such as~\cite{andersen2018nonintrusive, sharma2016data, karbasi2020non, zezario2020stoi}. These methods train a classification and regression tree or neural networks to predict intelligibility from features of noisy signals, therefore requiring a number of expensive human listening results or scores from intrusive predictors like STOI. Apart from the aforementioned methods, a series of works including FADE~\cite{schadler2015matrix}, and \cite{spille2018predicting, martinez2021prediction} leverage ASR models to predict speech reception thresholds (SRT), i.e., the signal-to-noise ratio (SNR) at which half of words within a group of noisy utterances are correctly recognised, rather than sequence-level intelligibility scores. The most recent work~\cite{martinez2021prediction} in the series does not require transcripts or reference signals for intelligibility prediction, while it uses identical noises for training and testing, and the recognition results need to be estimated.

\section{Method}
\label{sec:method}
In this section we describe how two sequence-level ASR uncertainty measures are formulated: \textit{confidence} and \textit{entropy}, with an ensemble method following the derivation in~\cite{malinin2020uncertainty}. The ensemble of models can be interpreted from a Bayesian perspective, i.e., regarding model parameters $\boldsymbol{\theta}$ as random variables and using a prior $\mathrm{p}(\boldsymbol{\theta})$ to compute the posterior $\mathrm{p}(\boldsymbol{\theta|\mathcal{D}})$ with a given dataset $\mathcal{D}$. As Bayesian inference is usually intractable for models like deep neural networks, it is possible to take advantage of an approximation  $\mathrm{q}(\boldsymbol{\theta})$ to $\mathrm{p}(\boldsymbol{\theta|\mathcal{D}})$ with a family of models with different parameters~\cite{hoffmann2021deep}. Monte-Carlo Dropout~\cite{gal2016dropout} and Deep Ensembles~\cite{lakshminarayanan2017simple} are two major approaches to generate ensembles, and the latter approach is exploited in this work.

\subsection{Uncertainty estimation}
Given the ASR training dataset containing variable-length sequences of input acoustic features $\{ x_{1}, \ldots, x_{N} \} = \boldsymbol{x} \in \mathcal{X}$, and the corresponding transcript targets $\{ y_{1}, \ldots, y_{L} \} = \boldsymbol{y} \in \mathcal{Y}$, an ensemble of $M$ ASR models $\{ \mathrm{P}(\boldsymbol{y}|\boldsymbol{x}; \boldsymbol{\theta}^{(m)})\}$ can be trained to achieve the approximated posterior $\mathrm{q}(\boldsymbol{\theta})$. The sequence-level predictive posterior $\mathrm{P}(\boldsymbol{y}|\boldsymbol{x}, \boldsymbol{\theta})$ can be computed as the expectation of the ensemble:
\begin{equation}
\begin{split}
\mathrm{P}(\boldsymbol{y}|\boldsymbol{x}, \boldsymbol{\theta}) &= \mathbb{E}_{\mathrm{q}(\boldsymbol{\theta})}[\mathrm{P}(\boldsymbol{y}|\boldsymbol{x}, \boldsymbol{\theta})] \\
&\approx \frac{1}{M} \sum_{m=1}^{M}\mathrm{P}\left(\boldsymbol{y}|\boldsymbol{x},\boldsymbol{\theta}^{(m)}\right),
\end{split}
\end{equation}
where $\boldsymbol{\theta}^{(m)} \sim \mathrm{q}(\boldsymbol{\theta}) \approx \mathrm{p}(\boldsymbol{\theta}|\mathcal{D})$. The sequence-level entropy $\mathrm{H}(\boldsymbol{y}|\boldsymbol{x}, \boldsymbol{\theta})$ can be expressed as:
\begin{equation}
\begin{split}
\mathrm{H}(\boldsymbol{y}|\boldsymbol{x}, \boldsymbol{\theta}) &= \mathbb{E}_{\mathrm{P}(\boldsymbol{y}|\boldsymbol{x}, \boldsymbol{\theta})}[-\ln \mathrm{P}(\boldsymbol{y}|\boldsymbol{x}, \boldsymbol{\theta})] \\
&=-\sum_{\boldsymbol{y} \in \mathcal{Y}} \mathrm{P}(\boldsymbol{y}|\boldsymbol{x}, \boldsymbol{\theta}) \ln \mathrm{P}(\boldsymbol{y}|\boldsymbol{x}, \boldsymbol{\theta}).
\end{split}
\end{equation}

It is usually not possible to compute the posterior $\mathrm{P}(\boldsymbol{y}|\boldsymbol{x}, \boldsymbol{\theta})$ as $\mathcal{Y}$ is an infinite set with variable-length transcript sequences. However, an autoregressive ASR model could factorise the posterior into a product of conditionals:
\begin{equation}
\mathrm{P}(\boldsymbol{y}|\boldsymbol{x}, \boldsymbol{\theta}) = 
\prod^{L}_{l=1} \mathrm{P}(y_{l}|\boldsymbol{y}_{<l}, \boldsymbol{x}; \boldsymbol{\theta}), y_{l} \in \{ \omega_{1}, \ldots, \omega_{K} \},
\end{equation}
where $\omega$ represents the byte-pair encoding (BPE) token, and $K$ is the size of BPE vocabulary.

\textit{Confidence} is usually considered as the maximum predicted probability, and the sequence-level confidence $\mathcal{C}_{S}$ in this work is regarded as a combination of token-level confidence. In order to make fair comparison of sequences with variable lengths, a length normalisation rate is used~\cite{cover1999elements}, and $\mathcal{C}_{S}$ is computed as:
\begin{equation}
\mathcal{C}_{S} = \exp \left[ \frac{1}{L} \ln \sum_{l=1}^{L} \max{\frac{1}{M} \sum_{m=1}^{M} \mathrm{P}(y_{l}|\boldsymbol{y}_{<l}, \boldsymbol{x}; \boldsymbol{\theta}^{(m)})} \right].   
\end{equation}

\textit{Entropy} computation is usually challenging as the expectations of $\boldsymbol{y}$ are practically intractable, i.e., there are $K^{L}$ possible candidates for a $L$-length sequence $y_{L}$, and a forward-pass inference needs to be conducted for each hypothesis $y$. Meanwhile, beam-search in ASR inference stage is able to provide high-quality hypotheses and can therefore be considered as an importance-sampling which yields hypotheses from high-probability space. By using $B$ top hypotheses within a beam, the approximated sequence-level entropy $\mathcal{H}_{S}$ with simple Monte-Carlo estimation can be computed as:
\begin{equation}
\begin{split}
& \mathcal{H}_{S} = - \sum^{B}_{b=1} \frac{\pi_{b}}{L^{(b)}}  \ln{\mathrm{P}(\boldsymbol{y}^{(b)}|\boldsymbol{x}, \boldsymbol{\theta})}, \\
& \pi_{b} = \frac{\exp{\frac{1}{T}} \ln{\mathrm{P}(\boldsymbol{y}^{(b)}|\boldsymbol{x}, \boldsymbol{\theta})}}{\sum^{B}_{k}\exp{\frac{1}{T}} \ln{\mathrm{P}(\boldsymbol{y}^{(k)}|\boldsymbol{x}, \boldsymbol{\theta})}},
\end{split}
\end{equation}
where a calibration temperature $T$ can be introduced to adjust the distribution of hypotheses, and:
\begin{equation}
\ln{\mathrm{P}(\boldsymbol{y}^{(b)}|\boldsymbol{x}, \boldsymbol{\theta})} = \sum^{L^{(b)}}_{l^{(b)}=1} \ln{\frac{1}{M} \sum_{m=1}^{M} \mathrm{P}(y_{l}^{(b)}|\boldsymbol{y}^{(b)}_{<l}, \boldsymbol{x}; \boldsymbol{\theta}^{(m)})}.
\end{equation}

\subsection{ASR posterior}
The ASR model used in this work is based on the transformer architecture~\cite{vaswani2017attention}, which has shown impressive results recently. The model consists of a convolutional neural network-based front-end, a transformer-based encoder, and a transformer-based decoder. A mechanism combining the Connectionist Temporal Classification\,(CTC) and attention-based sequence to sequence\,(seq2seq) is used for the optimisation~\cite{kim2017joint}. When estimating the uncertainty, the predictive posterior for each token is expressed as:
\begin{equation}
\begin{split}
\mathrm{P}(y_{l}|\boldsymbol{y}_{<l}, \boldsymbol{x}; \boldsymbol{\theta}^{(m)}) &= \lambda \mathrm{P}_{CTC}(y_{l}|\boldsymbol{y}_{<l}, \boldsymbol{x};\boldsymbol{\theta}^{(m)}) \\
&+ (1 - \lambda) \mathrm{P}_{seq2seq}(y_{l}|\boldsymbol{y}_{<l}, \boldsymbol{x}; \boldsymbol{\theta}^{(m)}),
\end{split}
\end{equation}
where $\lambda$ is a weighting coefficient.

\section{Experiments and Results}
\label{sec:experiments}
Our experiments are conducted on the Noisy Grid corpus~\cite{barker2007modelling} and the round one database of Clarity Prediction Challenge (CPC1)~\cite{barker2022the} which will be introduced in details in the next sub-sections. For both experiments, ensembles of six ASR models are employed to estimate uncertainty. As the entropy is supposed to be negatively correlated with intelligibility, negative entropy $-\mathcal{H}_{S}$ is used for evaluation. In addition, we evaluate word correctness score (WCS), defined as the number of words that are correctly recognised divided by the total number of words in the utterance, from ensembles of ASR models.

ASR models used in this work are all finetuned from the SpeechBrain~\cite{ravanelli2021speechbrain} released LibriSpeech model\footnote{huggingface.co/speechbrain/asr-transformer-transformerlm-\protect\\librispeech} with only different random seeds. The ASR models take 80-channel log mel-filter bank coefficients of an utterance with 16\,kHz sampling rate as input features. The convolutional front-end consists of three 2D convolutional layers, and the encoder and the decoder consists of twelve and six multi-head attention transformer blocks, respectively. The weighting coefficient $\alpha$ is set to 0.4 for uncertainty estimation. The calibration temperature $T$ is kept as 1. The top 10 hypotheses within the beam are used for entropy estimation.

Three metrics, including root mean square error\,(RMSE), normalised cross-correlation coefficient\,(NCC), and Kendall’s Tau coefficient\,(KT), are used to evaluate the correlation between the intelligibility scores from listening results, which are represented by the WCS, and the ASR WCS, the estimated uncertainty measures $\mathcal{C}_{S}$, $-\mathcal{H}_{S}$. Following the convention of evaluating intelligibility prediction, we report the correlations achieved by applying a logistic mapping function $f(x) = 1 / [1 + \exp(ax + b)]$, because RMSE and NCC could be invalid in non-linear cases, and the monotonicity correlation is already of great interest for analysis and optimisation. For the proposed method, the two parameters $a$ and $b$ are optimised in the development set with non-linear least squares\footnote{docs.scipy.org/doc/scipy/reference/generated/scipy.optimize.\protect\\curve\_fit.html}, and used in the test set to map the estimated uncertainty to the predicted intelligibility. For the baseline system, the parameters are optimised on the combined training and development sets.

\subsection{Noisy Grid corpus}
\label{subsec:ngrid}

\begin{figure}[t]
  \centering
  \includegraphics[width=0.9\linewidth]{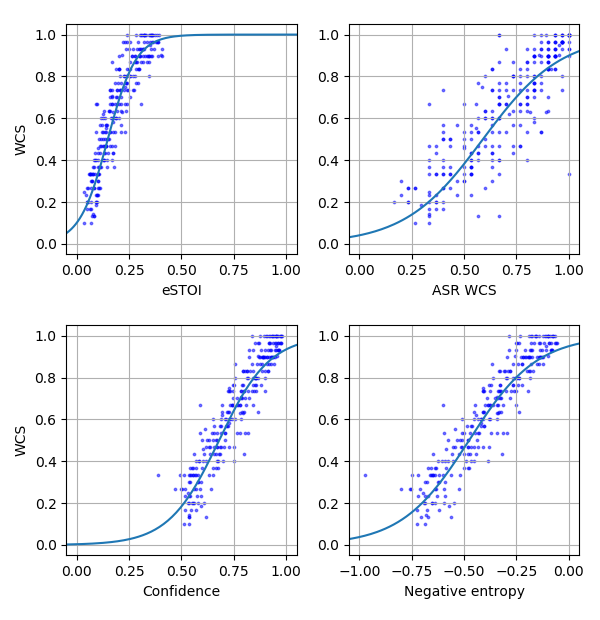}
  \caption{Predicted intelligibility measures on Noisy Grid Corpus test set, including eSTOI, and ASR WCS, confidence $\mathcal{C}_{S}$, negative entropy $-\mathcal{H}_{S}$, from the ensemble of ASR models optimised with the NGrid, versus the listening result WCS, in addition with the logistic mapping functions.}
  \label{fig:noisygrid}
\end{figure}

\subsubsection{Database}
The Noisy Grid corpus is an extension to the original Grid corpus~\cite{cooke2006audio} with added speech shaped noise\,(SSN) at 12 different SNR levels from -14\,dB to 40\,dB. Each Grid utterance consists of six words following the structure of ``command-color-preposition-letter-digit-adverb", and the words are randomly selected within a limited vocabulary of $[4, 4, 4, 25, 10, 4]$ words. The listeners are asked to identify ``color", ``letter", and ``digit" in the listening tests, therefore the WCS for each utterance can only be $[0, 1/3, 2/3, 1]$. In order to make the distribution of WCS relatively more continuous, the reported WCS is averaged over ten utterances at the same SNR level. The database comprises utterances spoken by 34 speakers, in which the utterances of 22 speakers are used as training set for ASR optimisation, 6 speakers as development set, and 6 speakers as test set. We observed that over $90\,\%$ utterances, whose SNRs are equal to or higher than 0\,dB, have perfect WCS in the listening tests. In order to even the distribution of the database, we report the results of utterances whose SNRs are lower than 0\,dB.

\subsubsection{Setup}
We exploit STOI and extended-STOI\,(eSTOI)~\cite{jensen2016algorithm}\footnote{https://github.com/mpariente/pystoi} as the baseline intelligibility predictors. Both STOI and eSTOI are intrusive measures taking advantage of the correlation between the acoustic features of clean reference signals and corresponding degraded processed signals. Because the Grid corpus has a limited vocabulary, the inference of the ASR models is strictly constrained within the Grid dictionary. As the ASR models operate at 16\,kHz, the Noisy Grid utterances are downsampled from 25\,kHz to 16\,kHz.

To investigate the impact of prior knowledge of the ASR models (the data used for ASR optimisation) could have on intelligibility prediction, we employ different ensembles of models including: ASRs finetuned on the training sets of (1) LibriSpeech\,(LS); (2) clean Grid corpus\,(CGrid); (3) clean Grid mixed with DEMAND noise~\cite{thiemann2013diverse} at SNRs from -15\,dB to 15\,dB\,(DGrid); (4) the original noisy Grid corpus\,(NGrid). The ensemble of ASR models finetuned on LS are optimised for two epochs, and those finetuned on CGrid, DGrid, NGrid are optimised for 10 epochs.

\begin{table}[t]
\centering
\caption{Correlation evaluation between the listening result WCS and predicted intelligibility measures on Noisy Grid Corpus test set.}
\resizebox{0.9\linewidth}{!}{
\begin{tabular}{l|c|c|c|c|c}
\toprule
& WER & Measure & RMSE $\downarrow$ & NCC $\uparrow$ & KT $\uparrow$\\\midrule
STOI & - & - & 0.154 & 0.853 & 0.670 \\ \midrule
eSTOI & - & - & 0.100 & 0.928 & 0.762 \\ \midrule
\multirow{3}{*}{LS} & \multirow{3}{*}{49.09}
& $\mathcal{C}_{S}$ & 0.172 & 0.762 & 0.572\\
& & $-\mathcal{H}_{S}$ & 0.166 & 0.788 & 0.595\\
& & WCS & 0.206 & 0.607 & 0.440\\ 
\midrule
\multirow{3}{*}{CGrid} & \multirow{3}{*}{32.88}
& $\mathcal{C}_{S}$ & 0.224 & 0.521 & 0.329\\
& & $-\mathcal{H}_{S}$ & 0.235 & 0.444 & 0.302\\
& & ASR WCS & 0.148 & 0.825 & 0.650\\ 
\midrule
\multirow{3}{*}{DGrid} & \multirow{3}{*}{21.03}
& $\mathcal{C}_{S}$ & 0.098 & 0.925 & 0.767\\
& & $-\mathcal{H}_{S}$ & 0.099 & 0.924 & 0.768\\ 
& & ASR WCS & 0.115 & 0.901 & 0.754\\
\midrule
\multirow{3}{*}{NGrid} & \multirow{3}{*}{17.04}
& $\mathcal{C}_{S}$ & \textbf{0.093} & \textbf{0.937} & 0.790 \\
& & $-\mathcal{H}_{S}$ & 0.094 & 0.936 & \textbf{0.791}\\ 
& & ASR WCS & 0.144 & 0.844 & 0.695\\
\bottomrule
\end{tabular}
}
\label{table:noisygrid}
\end{table}

\subsubsection{Results}
Table~\ref{table:noisygrid} lists the evaluation results on the Noisy Grid test set. Figure~\ref{fig:noisygrid} shows the eSTOI predicted intelligibility scores, ASR WCS, confidence and negative entropy from the ensemble of ASR models finetuned on NGrid versus the listening result WCS along with their logistic mapping functions. The result shows that the uncertainty estimated by the ensemble of ASR models optimised with NGrid is highly correlated with speech intelligibility and outperforms STOI, eSTOI. In addition, the uncertainty is better at intelligibility prediction than ASR WCS. The confidence is slightly more correlated with intelligibility than entropy in terms of RMSE and NCC, while the entropy performs slightly better in terms of KT. 

The word error rates (WER) of Noisy Grid test set for each ensemble of ASR models (which vary by their degree of prior knowledge of the evaluated ASR models) are also shown in Table~\ref{table:noisygrid}. It shows that a strong prior knowledge of the test data leads to a high correlation between ASR uncertainty and speech intelligibility based on the results of CGrid, DGrid, and NGrid. However, it can be observed that when the ASR models have no knowledge of the noisy signals, the confidence and negative entropy of LS finetuned ensemble could outperform the CGrid finetuned ensemble. It is also worth noting that ASR models optimised on DGrid, i.e., different type of noises from the Noisy Grid test set, could also produce competitive results.

\subsection{CPC1}
\label{subsec:cpc1}

\begin{figure}[t]
  \centering
  \includegraphics[width=0.9\linewidth]{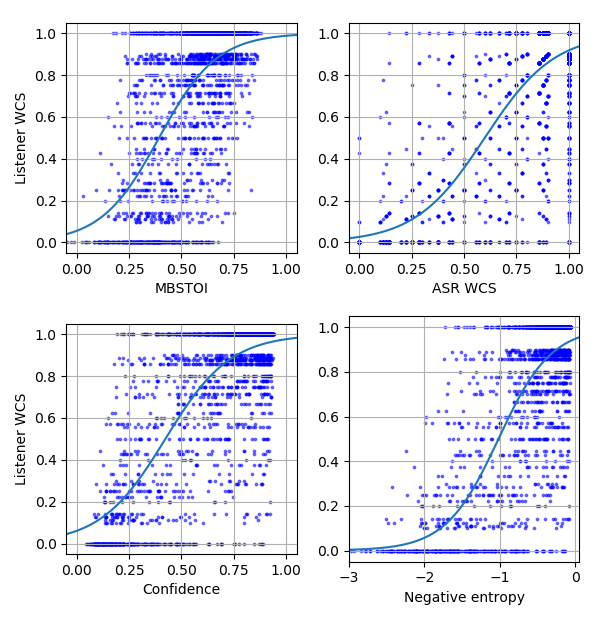}
  \caption{Predicted intelligibility measures on CPC1 closed evaluation set, including the baseline, ASR WCS, and confidence $\mathcal{C}_{S}$, negative entropy $-\mathcal{H}_{S}$, from the ensemble of MSBG+CLS+CPC1 ASR models,  versus the listening result WCS, in addition with the logistic mapping functions.}
  \label{fig:cpc1}
\end{figure}

\subsubsection{Database}
For the purpose of advancing hearing aid intelligibility prediction, the CPC1 database provides a large number of binaural signals and their corresponding responses made by hearing impaired listeners. Each signal corresponds to a noisy scene, which is simulated by mixing a target utterance and a segment of noise in a room, and enhanced by a machine learning hearing-aid system based on the listener's hearing loss measure. The complete database consists of 6 speakers, 10 hearing aid systems and 27 listeners. Two separate but related tracks are included in CPC1: (1) \textit{closed-set}, in which the evaluation hearing-aid systems and listeners are the same as those in the training data; (2) \textit{open-set}, in which the hearing-aid systems or listeners in the evaluation set are different from those in the training data. Readers are referred to \cite{barker2022the} for full details. In both tracks, the training/development scenes are split between 70~\% and 30~\%, and the results on the extra evaluation set are reported.

\subsubsection{Setup}
Since in CPC1 the listeners are hearing impaired and the signals are binaural, the CPC1 baseline system employs a combination of Cambridge MSBG hearing loss simulator~\cite{baer1993effects, baer1994effects, moore1993simulation, stone1999tolerable} and MBSTOI~\cite{andersen2018refinement}. The MSBG simulator applies simulated degradation to an input signal according to the hearing loss measures of a listener, and MBSTOI is a refined version of binaural STOI. 

To estimate the uncertainty of a binaural signal from an ensemble of ASR models, the signal is resampled to 16\,kHz after processing with the MSBG model. Uncertainty of the left and right channel of each binaural signal is estimated independently, and a better ear principle is applied for the binaural uncertainty, i.e., the higher value of $\mathcal{C}_{S}$ or $-\mathcal{H}_{S}$ is regarded as the binaural uncertainty. The same better ear rule is also applied to the left and right ASR WCS.

The pretrained ASR models are first finetuned on the LibriSpeech (LS) for two epochs. Furthermore, they are optimised with LS \textit{train-clean-100} added with noises from the training set in the first round Clarity Enhancement Challenge~\cite{graetzer2021clarity} for 10 epochs. Finally, the models are optimised with the CPC1 training set for another 10 epochs. Therefore, the ASR models possess knowledge of clean, noisy, and processed speech signals. For the \textit{closed-set} experiments, we trained two ensembles of ASR models with and without the MSBG hearing loss model processed signals.

\subsubsection{Results}
For the CPC1 \textit{closed-set}, the uncertainty estimated from the ensemble of ASR models are more strongly correlated with speech intelligibility than the baseline, and negative entropy gains a slight advantage over confidence. In terms of RMSE and NCC, the uncertainty also outperforms ASR WCS. On the contrary, ASR WCS performs better with regard to KT as WCS are discrete, i.e., \textit{tied} pairs are more likely to appear. In addition, the results show that the MSBG model could provide a slight advantage for intelligibility prediction.

The results on the \textit{open-set} are consistent with those on the \textit{closed-set}. It is also worth noting that, the baseline has a large performance drop as the evaluation signals are very different from the ones in the training set. However, the ASR models are quite robust to this mismatch as the WERs are similar, and achieve similar performances.

\begin{table}[t]
\centering
\caption{Correlation evaluation between the listening result WCS and predicted intelligibility measures on CPC1 evaluation set.}
\resizebox{\linewidth}{!}{
\begin{tabular}{l|c|c|c|c|c}
\toprule
& WER & Measure & RMSE $\downarrow$ & NCC $\uparrow$ & KT $\uparrow$\\\midrule
\multicolumn{6}{l}{\textit{Closed-set}} \\ \midrule
CPC1 Baseline & - & - & 0.285 & 0.621 & 0.398 \\ \midrule
Proposed  & \multirow{3}{*}{25.17}
& $\mathcal{C}_{S}$ & 0.241 & 0.751 & 0.472\\
\multirow{2}{*}{without MSBG}& & $-\mathcal{H}_{S}$ & 0.239 & 0.754 & 0.477\\ 
& &ASR  WCS & 0.249 & 0.730 & 0.525\\
\midrule
Proposed & \multirow{3}{*}{30.33}
& $\mathcal{C}_{S}$ & 0.234 & 0.767 & 0.497 \\
\multirow{2}{*}{with MSBG}& & $-\mathcal{H}_{S}$ & \textbf{0.233} & \textbf{0.768} & 0.499\\ 
& & ASR WCS & 0.249 & 0.731 & \textbf{0.526}\\ \midrule

\multicolumn{6}{l}{\textit{Open-set}} \\ \midrule
CPC1 Baseline & - & - & 0.365 & 0.529 & 0.391 \\ \midrule
Proposed & \multirow{3}{*}{30.93}
& $\mathcal{C}_{S}$ & 0.248 & 0.729 & 0.512 \\
\multirow{2}{*}{with MSBG}& & $-\mathcal{H}_{S}$ & \textbf{0.246} & \textbf{0.734} & 0.512\\ 
& & ASR WCS & 0.253 & 0.717 & \textbf{0.530}\\
\bottomrule
\end{tabular}
}
\label{table:cpc1closed}
\end{table}

\section{Conclusions}
\label{sec:conclusion}
In this paper, we have shown that the sequence-level uncertainty of DNN-based ASR models is strongly correlated with speech intelligibility. Therefore, the estimated confidence and entropy from an ensemble of ASR models can be used as effective non-intrusive intelligibility predictors. In addition, the uncertainty estimation is unsupervised requiring no explicit references, i.e., no listening WCS nor reference clean signals are needed for training the predictor. The experimental results on two databases show that the proposed method can outperform STOI and its variants, and is better than ASR WCS at intelligibility prediction.

\bibliographystyle{IEEEtran}
\bibliography{mybib}
\end{document}